\documentstyle[preprint,aps]{revtex}
\input epsf.sty
\newcommand{\half}{\mbox{$\textstyle \frac{1}{2}$}}

\begin{document}

\draft

\title{Quantum Channel Capacity of Very Noisy Channels}

\author{David~P.~DiVincenzo,$^1$ Peter~W.~Shor$^2$ and John A. Smolin$^1$}

\address{$^1$IBM T. J. Watson Research Center, Yorktown Heights, NY 10598}
\address{$^2$AT\&T Research, Murray Hill, NJ 07974}

\date{\today}
\maketitle
\begin{abstract}
We present a family of additive quantum error-correcting codes whose
capacities exceeds that of quantum random coding (hashing) for very
noisy channels.  These codes provide non-zero capacity in a
depolarizing channel for fidelity parameters $f$ when $f> .80944$.
Random coding has non-zero capacity only for $f>.81071$; by analogy to
the classical Shannon coding limit, this value had previously been
conjectured to be a lower bound.  We use the method introduced by Shor
and Smolin of concatenating a non-random (cat) code within a random
code to obtain good codes.  The cat code with block size five is shown
to be optimal for single concatenation.  The best known
multiple-concatenated code we found has a block size of 25.  We derive
a general relation between the capacity attainable by these
concatenation schemes and the coherent information of the inner code
states.

\end{abstract}

\narrowtext

\section{Introduction}
\label{introduction}

It still comes as a surprise to many physicists that the error
correction techniques that we now know to exist for quantum states are
basically digital and not analog.  Even after the discovery of quantum
error correcting codes\cite{SG}, it was felt that the analog metaphor
must be more appropriate; after all, a quantum state is specified by a
continuous set of complex numbers, and the fundamental physical
process which we consider as ``noise'' on the quantum state, unitary
transformations involving the state and its environment, also are
drawn from a continuous and not a discrete set.  Nevertheless, the
entangled structure of quantum states, which has no analogy in
classical mechanics, permits an essentially digital treatment
of errors.  

In fact, quantum error correction as we presently understand
it\cite{SG,purification,BDSW,lafknill,Schumacher1,Schumacher2,CS,Steane,Steane2,ekert,LF,Gotts,Cal1,SSorig,GF4,SL,bds97,cleve}
is required to be oblivious to the continuous nature of the quantum
state: error correction is accomplished by using a coded subspace such
that the effect of the errors and error-correction scheme are both
independent of the direction of the state vector in that subspace.
Furthermore, these error-correcting actions are not continuous, but
are drawn from a discrete set.  This is related to the fact that the
continuous action of the environment can also be ``digitized,'' in the
following sense: quite generally, noise on a quantum state can be
described as a transformation which takes a pure quantum state $\Psi$
to a mixed state $\rho$ given as an ensemble of pure states
$\{A_i\Psi\}$.  Each of the set of operators $A_i$ can be written as a
linear combination of some fixed operator basis,
$A_i=\sum_ja_{ij}E_j$.  The fixed set of operators $E_j$ are the
``error operators'' of the quantum channel.  There form a finite set
with $d^2$ elements, where $d$ is the dimension of the Hilbert space
of $\Psi$.  For qubits ($d=2$) it has become conventional to use the
error basis $E_j=I\ {\rm (identity)},\ \sigma_x,\ \sigma_y,\ \sigma_z$
(the Pauli matrices).

These four operators have a very ``digital'' interpretation.  In the
simplest memoryless noisy bitwise channel, the ``binary symmetric
channel'', each bit is either left alone or is flipped.  This
corresponds to the actions of the operators $I$ and $\sigma_x$ on the
$z$-basis qubit states.  In the quantum case there are just two other
actions; we refer to the action of $\sigma_z$ as a ``phase flip'', and
that of $\sigma_y$ as ``both a bit flip and a phase flip.''  An
alternative point of view is to represent the quantum state as a
two-bit object (cf. \cite{BDSW,Cal1}), 
or an object in a four-element field GF(4) \cite{GF4}, so
that the four operators are just the four possible digital noise
actions on them.  The only ``analog'' features left in the description of
the quantum channel are the continuous amplitudes $a_{ij}$; but these
play a very similar role in the quantum noisy channel to the bit-flip
probabilities in the classical digital channel.

In this paper we will concern ourselves with the quantum capacity
of a simple qubit channel, the {\em depolarizing} channel.  This
channel is completely characterized by one fidelity parameter $f$;
with probability $f$ the qubit passes through the channel undisturbed
($A_1=\sqrt{f}I$), while with equal probability $g=(1-f)/3$ the qubit
is subjected to a rotation by one of the three Pauli matrix operators
$\sigma_{x,y,z}$ ($A_{2,3,4}=\sqrt{g}\sigma_{x,y,z}$).

Defining the quantum capacity requires a discussion of the quantum
error correction codes mentioned above.  This discussion will be given
in detail in Sec. II.  Suffice it to say now that many quantum codes
$[n,k,d]$ are now known
\cite{SG,BDSW,lafknill,CS,Steane,Steane2,LF,Gotts,Cal1,SSorig,GF4}, in
which an arbitrary state of $k$ qubits are coded into a state of $n>k$
qubits in such a way that if no more than $t\equiv\lfloor d/2\rfloor$
of the $n$ qubits are subjected to an error, the original $k$-qubit
state can nevertheless be perfectly recovered.  The {\em rate} of this
code is $r=k/n$.

With this, the {\em quantum capacity} $Q(\chi)$ of a quantum channel
$\chi$ can be defined: 
$Q(\chi)$ is the maximum number $Q$ such that
for any rate $R<Q$ and any $\delta>0$ there exists a quantum code ${\cal C}$
with rate $k/n\geq R$
such that after the action of
$\chi$ any state $\psi$ encoded by $\cal C$ can be recovered 
with fidelity at least $1-\delta$ at the receiving end of the 
channel\cite{BDSW,bds97,cleve}.

Naively one might expect there to be a relationship between the
achievable capacity $Q$ for a depolarizing channel of fidelity $f$ and
the rate $r$ of a code in which $1-{t\over{n}}\approx f$, since $1-f$
is the expected fraction of qubits on which errors will occur.  In
fact, there is no direct relationship because the definition
of capacity does not require that all errors of weight less
than $t$ are correctable, but only that the fraction
of uncorrectable errors vanishes for large block sizes.
For example, Rains has shown\cite{Rains}
that all families of codes for which $1-\frac{t}{n}<{5\over 6}$ have
vanishing rate $r$; nevertheless, $Q$ is known to be greater than zero
for a range of $f<{5\over 6}$.  Indeed, in our previous work on
``one-way hashing,'' \cite{purification,BDSW} 
we identified a method for which a non-zero
capacity could be attained down to about $f=.81$ \cite{micro1}.  
The capacity attained has the form of one minus a von-Neumann entropy (see
Eq.~(\ref{hash})).  This quantum expression bears a close resemblance
to the result of classical information theory, where the maximum
information reliably transferable though a noisy channel is limited by
the Shannon bound \cite{shannon}
\begin{equation} 
C\le 1-H(\chi) 
\end{equation}
where $H(\chi)$ is the average entropy introduced in a bit by
classical channel $\chi$.  

In the classical problem the Shannon bound is achieved by a random
coding procedure; the one-way hashing protocol which we
invented is the natural quantum analog of random coding.  Thus, it was
natural to expect that Eq.~(\ref{hash}) would also be the upper
bound on the quantum capacity for the depolarizing channel.

However, the quantum coding problem has not proved to be exactly
parallel to the classical one; recent work has identified several
important properties (the pipelining inequality, and subadditivity of
mutual information) \cite{Barnum} which are true for classical
capacity measures, but are not for the quantum version.  More
concretely, it was recognized early that quantum error correcting
codes can have a property referred to as ``degeneracy'' which is not
permitted in the quantum case: two different errors may be
indistinguishable by their error syndromes, but may nevertheless be
both correctable (see Sec. \ref{conclusions}).  Spurred by intuitive
ideas of how this degeneracy might improve the capacity of the quantum
channel, Shor and Smolin \cite{SSorig} explored some {\em non-}random
coding strategies, and found a range of depolarizing channels (very
noisy ones) for which the obvious analog of the Shannon bound is
violated; a higher capacity is attained than for random codes.

The main point of this paper is to present the Shor-Smolin discovery
using the more modern and streamlined tools for describing quantum
coding which have been developed recently.  We will formulate the
capacity calculation in terms of the orthogonal-group formalism 
\cite{Gotts,Cal1} which 
has proved very successful in systematizing almost all known quantum
codes (the additive codes) \cite{nonad}.  
We identify new quantum weight enumerators (cf. \cite{SL}), for the stabilizer {\em
cosets}, with which a compact expression for the capacity can be
given.  We show that the Shor-Smolin codes can be understood in the
language of code {\em concatenation} which has been very popular for
the discussion of fault-tolerant quantum computation
\cite{Shorfault,DS,Aharonov,knillfault,Preskill}.  
We establish a
rather general relation between concatenated-code capacities and the
quantum {\em coherent information} \cite{Schumacher1,Schumacher2},
a quantity believed to be (but not
proved to be in general) the quantum capacity \cite{Lloyd,Barnum}.
Finally, we show that, using the 
coset-enumerator formulation, closed-form expressions can be derived
for the capacities of the original Shor-Smolin protocols, which
permit us to perform a more extensive quantitative exploration of the
performance of these codes.  We do not yet know what the actual
attainable capacity of the depolarizing channel is, but hopefully
the techniques explored here may provide a clue of how to obtain
this result.

This paper is organized as follows: Sec. \ref{sec2} briefly reviews the 
orthogonal-geometry group theory which has been introduced for
the classification of quantum codes, and introduces the necessary 
coset weight polynomials.  Sec. \ref{sec3} derives the average-entropy
expression for concatenated codes as originally obtained by
Shor and Smolin.  Sec. \ref{sec4} shows that the capacity attained
by the concatenation procedure is equal to the quantum coherent
information.  Sec. \ref{sec5} presents the compact expressions which we
have obtained for the concatenation using the ``cat'' code, which
give capacities exceeding the random-coding bound.   
Sec. \ref{conclusions} presents our conclusions, and
some thoughts about the use of degeneracies to attain improved
capacities using concatenated codes.

\section{Group-Theoretic Characterization of Codes}
\label{sec2}

We consider the group $\bar E$ introduced in \cite{Cal1} which
describes all possible standard errors on $n$ uses of the channel
(described by products of Pauli-matrix operators on a set of $n$
qubits).  The bar indicates that the group is understood to be taken
modulo phases $\pm1$, $\pm i$.  The dimension of $\bar E$ is $2^{2n}$.
As \cite{Cal1} showed, the Abelian subgroups of $E$ play a central
role in the theory of quantum error correcting codes.  Consider such
an Abelian subgroup $S$; again, we will work only with $\bar S$, from
which phases have been removed.  We define $k<n$ by specifying that
$\bar S$ has $n-k$ generators (and thus is of dimension $2^{n-k}$);
then any of the $2^k$-dimensional eigenspaces (denoted ${\cal
C}_i,~0\le i<2^k$) of this set of Pauli-matrix operators forms a
quantum code.  We now introduce $S^\perp$ and $\bar S^\perp$ where
$S^\perp$ consists of all elements in $E$ which
commute with all the elements of $S$ and $\bar S^\perp$ is $S^\perp$ modulo
phases.  The dimension of the set
$\bar S^\perp$ is $2^{n+k}$.  To analyze the error-correction
capability of the code, we define the weight of an operator $e$ in
$E$, ${\rm wt}(e)$, as the number of Pauli-matrix operators (either
$\sigma_x$, $\sigma_y$, or $\sigma_z$) appearing in $e$.  If the
minimal-weight element of $\bar S^\perp\backslash\bar S$ (i.e., the set
$\bar S^\perp$ excluding the elements in $\bar S$) has weight $d$,
then the correct state can be restored after $d-1$ erasures, which
means also that it can correct arbitrary errors on any $t=\lfloor
d/2\rfloor$ qubits.  $d$ is referred to as the ``distance'' of the
quantum code, and the notation for the code is $[n,k,d]$.

All these facts have been discussed previously for quantum codes; but
for the present purposes we need to introduce some additional
mathematical objects, the {\em cosets} of $\bar S$ in $\bar E$.  To
understand why these cosets might be natural objects to consider for
quantum error correcting codes, we recall that the elements of $\bar
S$ are the {\em stabilizers} of the code; this means that if the error
suffered by the set of $n$ qubits is any member of $\bar S$, then the
code state is unaffected.  Note that in the same way any element of
the coset $\bar s_\alpha\bar S$ acts identically on the code; all such
elements act as if just the error $\bar s_\alpha$ had occurred.  For
this reason, these cosets will play a central role in the analysis
below.  Note that because $\bar E$ is Abelian (N.B. $E$ is {\em not}
Abelian) no distinction need be made between left and right cosets.

\begin{figure}
\epsfxsize=15cm
\leavevmode
\epsfbox{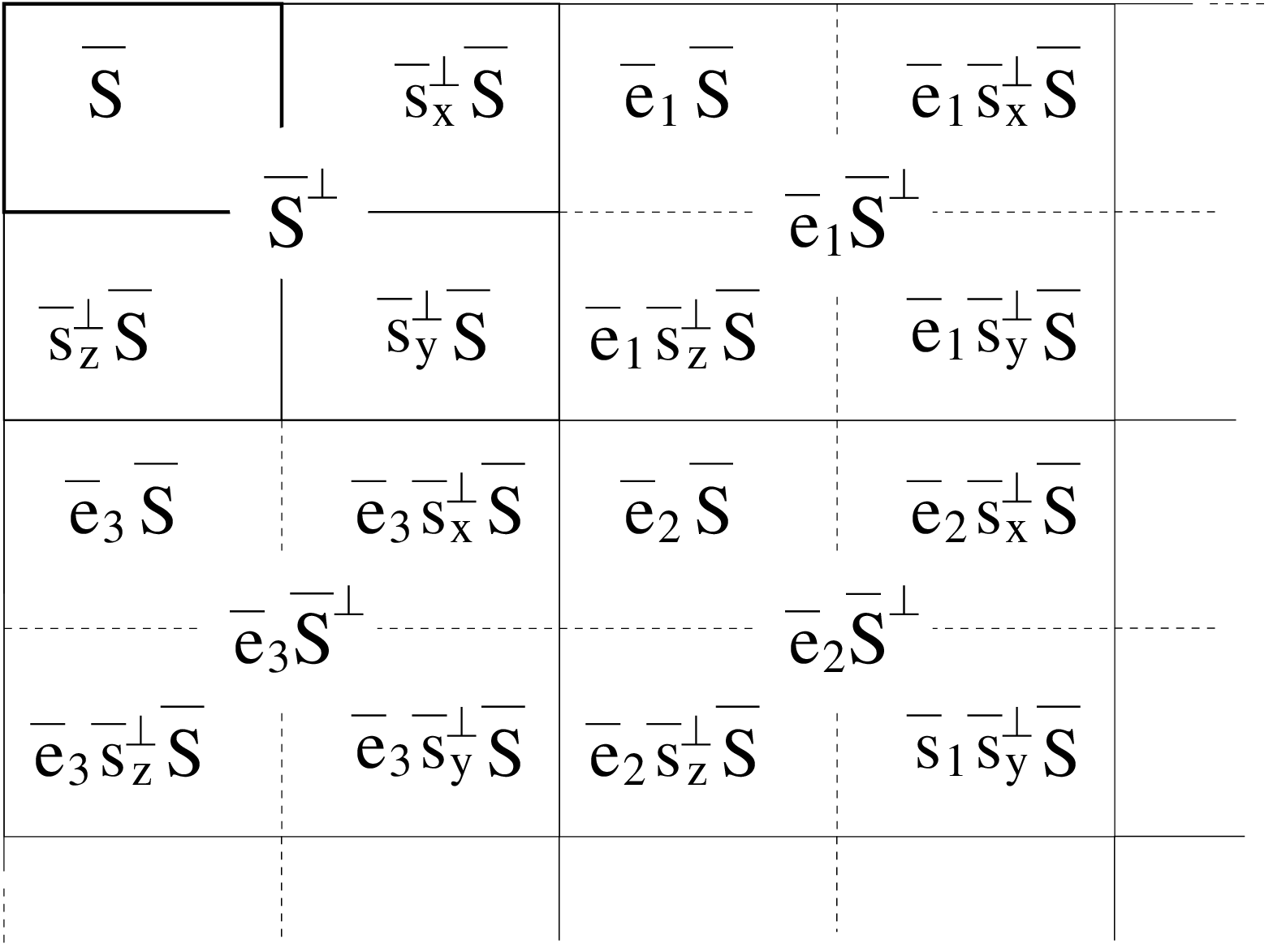}
\caption{Hierarchical partitioning of the set $\bar E$ into cosets of
$\bar S^\perp$, and those in turn into cosets of $\bar S$.  The case
of $\bar S^\perp$ dividing into four cosets is special to the case of
coding a single qubit.
\label{fig1}}
\end{figure}

We will need to consider three different coset partitionings: 1) The
cosets of $\bar S^\perp$ in $\bar E$.  Consider the ``transversal'' of
$\bar E$, the set $G=\{\bar\alpha\}\subset\bar E$ 
which generates the coset decomposition of
$\bar S^\perp$ in $\bar E$.  That is, $\bar e_\alpha\bar
S^\perp\bigcap\bar e_\beta\bar S^\perp=\emptyset$ if $\alpha\ne
\beta$, and $\bigcup_\alpha\bar e_\alpha\bar S^\perp=\bar E$.  The
dimension of $G$ is $2^{n-k}$.  We indicate the $\alpha$th coset of
the coset decomposition as $\bar e_\alpha\bar S^\perp$.  2) The cosets
of $\bar S$ in $\bar S^\perp$.  In this case we denote the set
generating all the distinct cosets as $G^\perp=\{\bar
s^\perp_\alpha\}\subset\bar E$.  The dimension of this set is
$2^{2k}$.  A typical coset is indicated as $\bar s^\perp_\alpha \bar
S$.  3) The cosets of $\bar S$ in $\bar E$.  This is just the direct
product of 1) and 2), the generating set is $G\otimes G^\perp$, and
the $(\alpha,\beta)$ coset is denoted $\bar e_\alpha\bar
s^\perp_\beta\bar S$.  This hierarchy of coset decompositions is
indicated in Fig.~\ref{fig1} for $k=1$.

We introduce a weight-enumerator polynomial $P$ for set $C$ as:
\begin{equation}
P(C)\equiv\sum_{e\in C} f^{n-{\rm wt}(e)}g^{{\rm wt}(e)},\label{pofc}
\end{equation}
where $C\subset\bar E$.

We note that this weight polynomial is directly related to the
Shor-Laflamme weight enumerator\cite{SL}.  Their weight function $A_d$ is
(we use the normalization choice of Rains\cite{Shad})
\begin{equation}
A_d({\cal O}_1,{\cal O}_2)=
\sum_{e'\in\bar E~|\atop{\rm wt}(e')=d}{\rm tr}(e'{\cal O}_1){\rm tr}
(e'{\cal O}_2).
\end{equation}
and their weight-enumerator polynomial is 
\begin{equation}
A(z,{\cal O}_1,{\cal O}_2)=\sum_{d=0}^nA_d({\cal O}_1,{\cal O}_2)z^d.
\end{equation}
To relate these to Eq.~(\ref{pofc}), note that
\begin{equation}
A_d(e,e)=2^{2n}\delta_{{\rm wt}(e),d},
\end{equation}
from which we see that
\begin{equation}
P(C)\equiv\sum_{e\in C}\sum_{d=0}^nA_d(e,e)f^{n-d}g^d=
\frac{f^n}{2^{2n}}\sum_{e\in C}A\left(\frac{g}{f},e,e\right)=
\left(\frac{f}{4}\right)^n\sum_{e\in C}A\left(\frac{1-f}{3f},e,e\right).
\end{equation}
So our new polynomials are simple functions of those which have been
introduced in previous work in quantum error correcting codes.

Our weight polynomial $P(C)$ has particular significance for the
depolarizing channel with fidelity f when $C$ is the one of the
various cosets which we introduced above:
\begin{itemize}
\item $P(\bar S)$ is the probability that the coded quantum state will
leave the depolarizing channel without error.  Each distinct operator
$e\in\bar S$ is an action of the channel which has this property, and
$f^{n-{\rm wt}(e)}g^{{\rm wt}(e)}$ is the probability of that action.
\item $P(\bar s_\alpha^\perp\bar S)$ gives the probability that the
coded state leaving the quantum channel is {\em detected} to have no
error, but has actually been rotated inside the code eigenspace by
$\bar s_\alpha^\perp$.  This is so from the definition of 
$\bar s_\alpha^\perp$: since it commutes with all elements of $S$,
it does not change the eigenvalues of $\bar S$ which are detected in
the channel-decoding operation\cite{DS}; and, since by definition 
$\bar s_\alpha^\perp\notin\bar S$, it performs a non-identity rotation
of the coded state inside the code eigenspace.  In fact, every member
of the coset $\bar s_\alpha^\perp\bar S$ performs the same rotation.
\item $P(\bar S^\perp)$ gives the probability that no error will be
detected upon decoding, regardless of whether the final quantum state
is correct or not.
\item $P(\bar e_\alpha\bar S^\perp)$ is the probability that decoding
detects error $\bar e_\alpha$, regardless of the rotation of the coded
state.
\item $P(\bar e_\beta\bar s_\alpha^\perp\bar S)$ is the probability
that decoding detects error $\bar e_\beta$, and the coded state is
rotated by $\bar e_\beta\bar s_\alpha^\perp$.
\end{itemize}
These will be the essential tools for developing a compact formula for
the attainable capacity for code states, and establishing the identity
of the coherent information with the Shor-Smolin quantum channel
capacity.

\section{Shor-Smolin concatenation procedure}
\label{sec3}

\begin{figure}
\epsfxsize=15cm
\leavevmode
\epsfbox{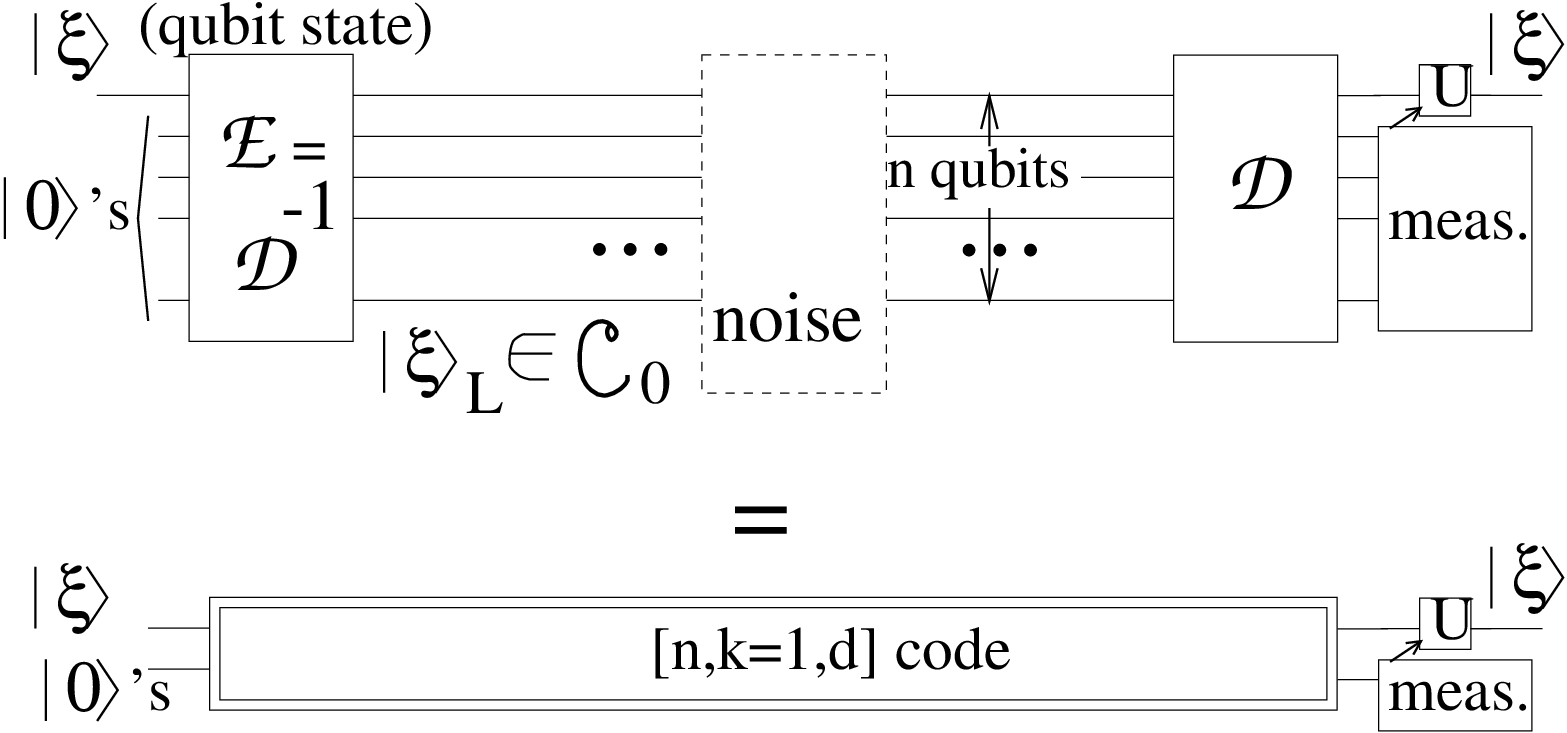}
\caption{Quantum channel coding, in which (top) the state to be
transmitted $|\xi\rangle$ is encoded by $\cal E$, transmitted through
the noisy channel, decoded by $\cal D$ and restored by $U$ after
syndrome measurement.  The entire encode-transmit-decode process can
be through of a module (double box, below) to be used in concatenation
(see Fig. \protect\ref{fig4}).
\label{fig2}}
\end{figure}

In order to formulate the main result, we first review the Shor-Smolin
procedure\cite{SSorig} for sending reliable qubit states, with a
finite capacity, over a depolarizing channel.  Just as in conventional
channel coding, it involves an additive code specified above by $\bar
S$.  In conventional channel coding shown in Fig. \ref{fig2}, the
additive code is used as follows: the state $|\xi\rangle$ to be
transmitted (we specialize in the figure to a single qubit state) is
rotated by the encoding unitary transformation $\cal E$ into the
eigenspace ${\cal C}_0$ of the operators in $\bar S$.  When this state
passes through the depolarizing channel, it is rotated into one of the
other eigenspaces ${\cal C}_m$ with some probability: we will analyze
this process in detail later.  Then after passage through the noisy
channel, the decoding transformation $\bar D$ places $n-k$ of the
qubits (the lower $n-1$ in the figure) in a state such that, when they
are measured in the standard basis, they give the eigenvalue of each
of the $n-k$ generators of $\bar S$, that is, it determines which of
the spaces ${\cal C}_m$ the state had been placed into by the
noise\cite{DS}.  So long as the errors produced by the channel are
restricted to have weight no greater than $\lfloor d/2\rfloor$, then a
rotation $U$ can always be determined which restores the state to its
noiseless form $|\xi\rangle$.

\begin{figure}
\epsfxsize=15cm
\leavevmode
\epsfbox{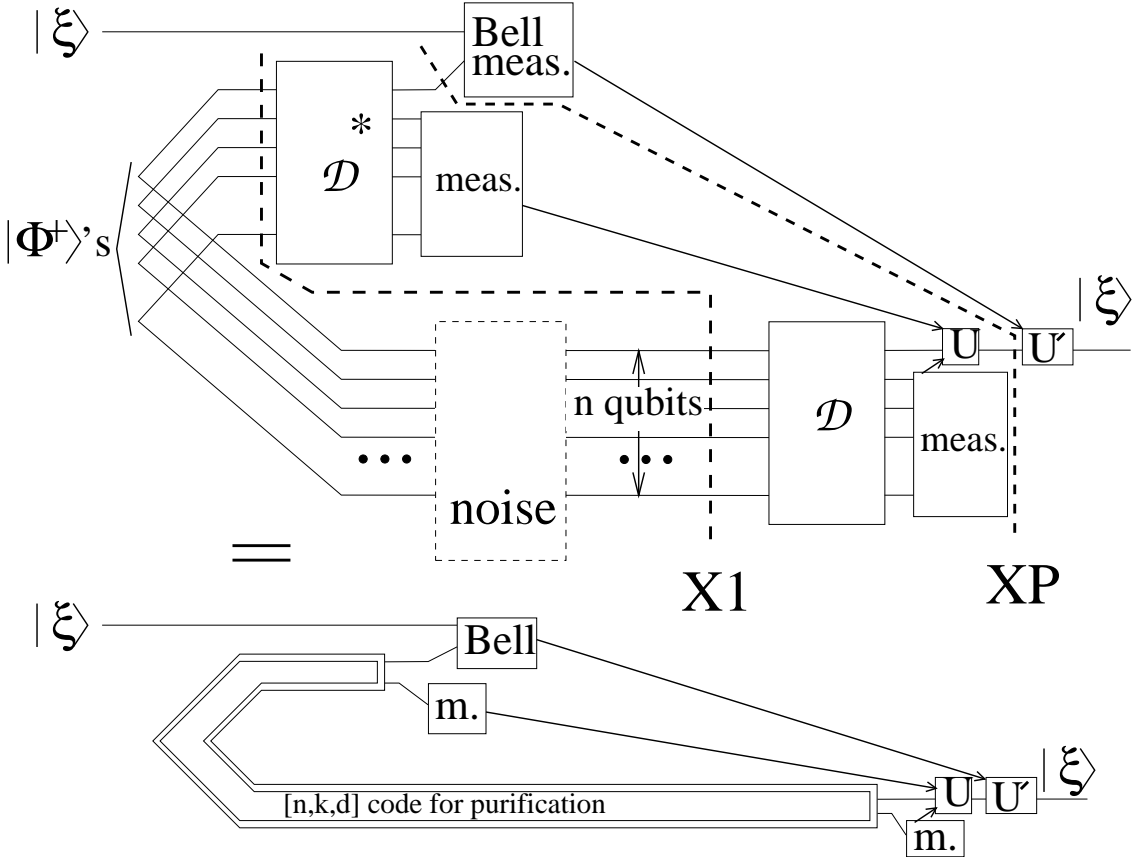}
\caption{A protocol for transmitting through the noisy channel
equivalent to Fig. \protect\ref{fig2} which uses entanglement
purification and teleportation.  The sender passes halves of Bell
states ($\Phi^+$) through the channel to the receiver; the degraded
pairs are purified (after slice $X1$) by $\cal D$ and ${\cal D}^*$
(same $\cal D$ as in Fig. \protect\ref{fig2}).  The purified pairs at
slice $XP$ can then be used to transmit the state $|\xi\rangle$ from
sender to receiver by teleportation.  Below: the Bell-state
distribution and processing with $\cal D$ and ${\cal D}^*$ may be used
as a module (double box) for the concatenation of
Fig. \protect\ref{fig5}.
\label{fig3}}
\end{figure}

The discussion below uses another protocol for coding shown in
Fig. \ref{fig3}, the purification protocol of
\cite{purification,BDSW}.  For the depolarizing channel the two
procedures of Figs.~\ref{fig2} and \ref{fig3} are completely
equivalent.  Ref.~\cite{BDSW} gives a detailed derivation of the
mapping of the first protocol to the second.  In the protocol of
Fig. \ref{fig3}, the sender begins with $n$ completely entangled
states, in this example the Bell state
\begin{equation}
\Phi^+={1\over{\sqrt{2}}}(|00\rangle+|11\rangle).
\end{equation}
The sender keeps one half of each of the $n$ Bell states, and the
other $n$ particles are sent through the depolarizing channel to the
receiver. When we are sending halves of EPR-Bell particles through the
channel, we no longer discuss the action of the channel in terms of
rotations among different code spaces (no coding transformation has
yet been applied to these states); rather, the state of the system at
slice $X1$ in Fig. \ref{fig3} (which in general is at two different
times for the two different sets of particles) is one in which the set
of Bell states has been rotated to a set of some of the other Bell
states with various probabilities which we will discuss shortly. The
full set of Bell states is
\begin{eqnarray}
\Phi^\pm={1\over{\sqrt{2}}}(|00\rangle\pm|11\rangle),\\
\Psi^\pm={1\over{\sqrt{2}}}(|01\rangle\pm|10\rangle).
\end{eqnarray}
The probability of a particular set of $n$ Bell states at slice $X1$ is
determined by the rule that the Bell state remains a $\Phi^+$ with
probability $f$, and becomes one of the three other states $\Phi^-$,
$\Psi^\pm$ with probability $g=(1-f)/3$.

Using the decoding transformations $\cal D$ and ${\cal D}^*$ in
Fig.~\ref{fig3}, followed by measurements on both ends, classical communication
from the sender to the receiver, and the final unitary transformation
$U$, the sender and receiver can come into possession of a
``purified'' $\Phi^+$ pair, which is then used to send the qubit state
$|\xi\rangle$ by teleportation \cite{teleportation}
(for details see \cite{BDSW}).

The two methods of employing the channel shown in Figs.~\ref{fig2} and
\ref{fig3} are completely equivalent.  But it will be useful to use
both points of view for explaining the generalized channel
transmission protocol of Shor and Smolin \cite{SSorig}, 
and we will continue our review using both languages.

We will need to apply our capacity definition of
Sec. \ref{introduction} to the purification picture.  The fidelity
$F^{\cal D}$ of the depolarizing channel output can be most simply be
defined in this picture in the following way: at the end of
purification (slice $XP$ in Fig. \ref{fig3}) the output is desired to
be a collection of $k$ $\Phi^+$ states; if the code scheme is a
successful one, then the overlap between the actual state at this
slice $\rho_{XP}$ and the desired Bell state will be high; thus the
fidelity for an encoding $\cal D$ is
\begin{equation}
F^{\cal D}=\langle(\Phi^+)^k|\rho_{XP}|(\Phi^+)^k\rangle.
\end{equation}
The capacity $Q$ is simply the best rate $k/n$ for a $\cal D$ for which 
this fidelity approaches unity, since each high-fidelity EPR pair
can be used to teleport one qubit.

The maximization of $Q$ has proved to be difficult.
But a variety of code families have been introduced for which finite
$Q$s are known, establishing useful lower bounds on the attainable
capacity.  One of the most useful is the sequence of {\em random
additive codes}, referred to in the original papers
\cite{purification,BDSW} as 
``one-way hashing.''  As the name suggests, these sequences
are built by selecting, at random, an Abelian subgroup $S$ from the
group of all Pauli matrices $E$ for successively larger block sizes
$n$.  Bennett {\em et al.}\cite{BDSW} show that almost all such
sequences attain the ``hashing capacity''
\begin{equation}
Q_H\equiv \lim_{n\rightarrow\infty}\frac{k}{n}=1-S_W(f).\label{hash}
\end{equation}
Thus, $Q_H(f)$ is a lower bound on the attainable capacity.  $S_W(f)$
is the von-Neumann entropy of one Bell state after one of its
particles has been passed through the depolarizing channel, and it is
given by
\begin{equation}
S_W(f)=-f\log f-3g\log g.
\end{equation}

Here is a brief explanation of why one-way hashing achieves the
capacity of Eq.~(\ref{hash}).  The entropy of the mixture of Bell
states at slice $X1$ is just $nS_W(f)$.  The decoding can be simply
thought of as a sequence of measurements of the $n-k$ operators which
are the generators of $\bar S$.  Each of these measurements has two
outcomes, splitting the set of possible remaining states in two; thus,
it has the potential for reducing the entropy of the state by one bit.
Ref. \cite{BDSW} provides arguments for why, for almost all choices of
$\bar S$ and for large $n$, each measurement in fact succeeds in
extracting one bit of entropy.  The total state remains a mixture of
Bell states, so that if $k$ is chosen so that the entropy is reduced
to zero, i.e., if $nS(W)-(n-k)=0$, then the Bell mixture becomes a
pure state, which is to say that the final state is one particular set
of known Bell states, which can always be rotated with $U$ to become a
set of $\Phi^+$ states.  Thus, purification has succeeded, and the
ratio $k/n$ attains the value given in Eq.~(\ref{hash}).

\begin{figure}
\epsfxsize=15cm
\leavevmode
\epsfbox{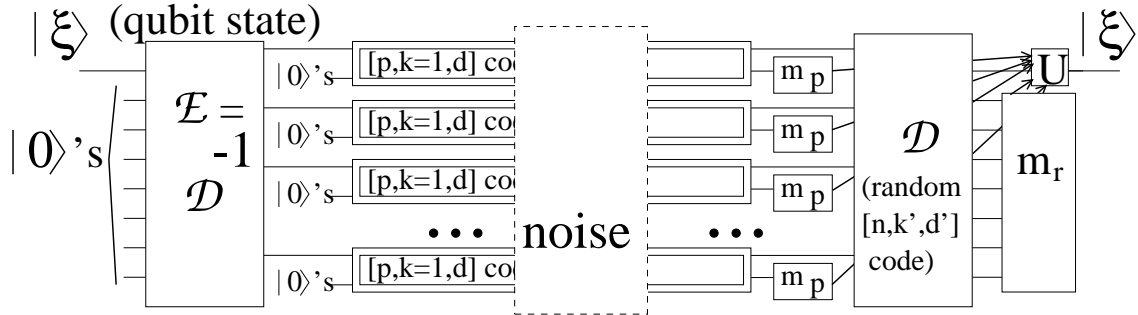}
\caption{Concatented coding for channel transmission.  The inner code (double
box) is the encode-transmit-decode module of Fig. \protect\ref{fig2}.  In the
Shor-Smolin procedure the outer part is a random code.
\label{fig4}}
\end{figure}

\begin{figure}
\epsfxsize=15cm
\leavevmode
\epsfbox{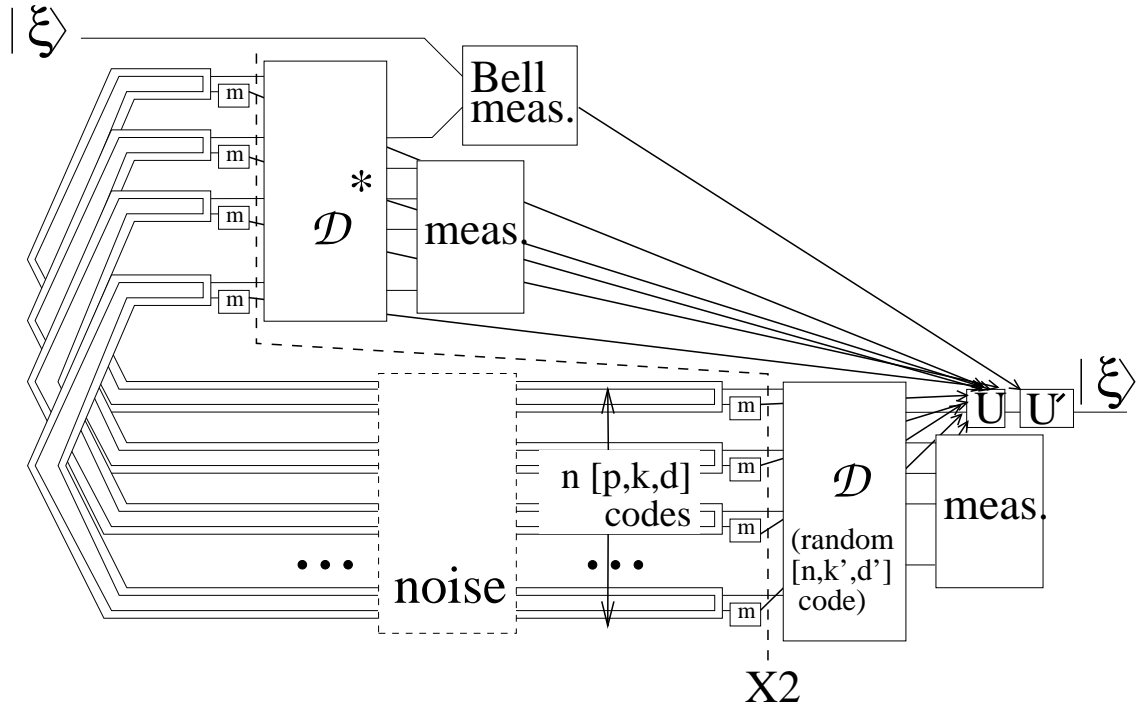}
\caption{Same as Fig. \protect\ref{fig4}, but using the
purification-teleportation protocol.
\label{fig5}}
\end{figure}

This result naturally raises the question of whether there exist any
{\em non-randomly} chosen sequences of codes which could attain a
capacity exceeding Eq.~(\ref{hash}).  While appeal to analogous
classical results and other thinking suggested that random coding
would be optimal, the Shor-Smolin construction which we now review
shows that higher capacities are attainable.  Their construction
involves what is known as concatenation; it is illustrated, for both
versions of the quantum coding protocols, in Figs. \ref{fig4} and
\ref{fig5}.  In the language of Fig. \ref{fig4}, the idea is that
instead of sending the qubits as encoded by the random encoder $\cal
E$ directly into the channel, they are encoded once again in another
additive code $[p,k,d]$, and it is these $n\times p$ qubits that are
finally sent through the channel.  The codes whose capacity we will
consider involve $n\rightarrow\infty$, but fixed $p$.  While we tend
to associate ``good'' (i.e., high-capacity) codes with large distance
$d$, we will find that the desirable inner $[p,k,d]$ codes actually
have a {\em small} distance $d$.  As we discuss at the end, it may
be the ``degeneracy'' of this code which is relevant.

Shor and Smolin showed \cite{SSorig} that the following capacity is
attainable by this concatenated scheme:
\begin{equation}
Q_{SS}=\frac{1}{p}(1-S_{X2}).
\end{equation}
The $1/p$ just comes from the fact that the whole scheme requires
$p\times n$ bits rather than just $n$ bits to be sent through the
channel.  $S_{X2}$ is the average entropy of each bipartite state at slice
$X2$ in Fig. \ref{fig5} (the total entropy at slice $X2$ is
$nS_{X2}$).  Shor and Smolin noted that this entropy is {\em not}
given by the von-Neumann entropy of the quantum state at this slice,
because of the presence of the results of the classical measurements.
Rather it is the {\em average} of the von-Neumann entropies of the
quantum states conditional on the measurement outcomes:
\begin{equation}
S_{X2}=\sum_{{i\in{\rm meas.}}\atop{\rm outcomes}}{\rm Pr}(i)S(\rho|i)=
\sum_{{i\in{\rm meas.}}\atop{\rm outcomes}}{\rm Pr}(i)h_4(\{{\rm Pr}
(B_j|i)\}).\label{SSx}
\end{equation}
It is this entropy that is to be reduced to zero by the random-hashing
stage of the decoding.  In the second part of Eq.~(\ref{SSx}) we have
specialized to the case where the inner code has $k=1$ (and thus
produces just one qubit-pair state in Fig. \ref{fig5}).  In this case the
output is a mixture of the four Bell states
$\{B_j\}=\Phi^\pm,~\Psi^\pm$, so that the entropy just involves the
probability of Bell state $B_j$ conditional on the particular
measurement outcome $i$:
\begin{equation}
{\rm Pr}(B_j|i)=\frac{{\rm Pr}(B_j,i)}{{\rm Pr}(i)},~~{\rm Pr}(i)=
\sum_{j=1}^4{\rm Pr}(B_j,i).\label{Prs}
\end{equation}
The $h_4$ function in Eq.~(\ref{SSx}) on the set $\{x_i\}$ is defined
by
\begin{equation}
h_n(\{x_j\})\equiv -\sum_{j=1}^nx_j\log_2x_j,~~~\sum_{j=1}^nx_j=1.
\label{hdef}
\end{equation}
By using the elementary algebraic properties of the $h_n$ function
$S_{X2}$ may be simplified so that $Q_{SS}$ is expressed as
\begin{equation}
Q_{SS}={1\over p}[1+h_N(\{{\rm Pr}(i)\})-h_{4N}(\{{\rm Pr}(B_j,i)\})].
\label{QSSsimp}
\end{equation}
Here $N$ is the number of distinct measurement outcomes;
for an additive $[p,k=1,d]$ code, $N=2^{p-k}=2^{p-1}$.  

The probabilities appearing in Eq.~(\ref{QSSsimp}) have appeared above;
in fact they are equal to
\begin{equation}
{\rm Pr}(i)=P({\bar e}_i{\bar S}^\perp),\label{rel1}
\end{equation}
\begin{equation}
{\rm Pr}(B_j,i)=P({\bar s}_j^\perp{\bar e}_i{\bar S}).\label{rel2}
\end{equation}
Eq.~(\ref{rel1}) follows from the fact that the members of the set
${\bar S}^\perp$ are, by definition, those errors which all lead to
the measurement which indicates the ``no-error'' condition; thus, its
cosets in $\bar E$, ${\bar e}_i{\bar S}^\perp$, each contain the
errors which all lead to the same measurement $i$.  Finally, the
weight polynomials are, as discussed above, constructed so as to
enumerate properly the probabilities of these sets.  Eq.~(\ref{rel2})
follows similarly: The set $\bar S$ indicates those errors which lead
to the ``no-error'' measurement {\em and} leave the Bell state $B_i$
in the correct $\Phi^+$ state.  Furthermore, the coset ${\bar
s}_j^\perp{\bar e}_i{\bar S}$ contains those errors which lead to
measurement $i$ and Bell state $B_j$.  It should be noted that the
error operations ${\bar s}_j^\perp$ have the effect of performing a
unitary operation of the coded qubit; the four operations are 1)
${\bar s}_0^\perp=I$ (the identity), which leaves the Bell state
$\Phi^+$ unaffected, 2) ${\bar s}_x^\perp$, which performs a coded
$\sigma_x$, leading to a final Bell state $B_x=\Psi^+$, 3) ${\bar
s}_y^\perp$ which performs $\sigma_y$ and leads to $B_y=\Psi^-$, and
4) ${\bar s}_z^\perp$ which performs $\sigma_z$ and leads to
$B_z=\Phi^-$.  So, the weight polynomial in Eq.~(\ref{rel2}) is
constructed to evaluate the probability that a member of the coset
occurs.

Finally we may rewrite the capacity equation as
\begin{equation}
Q_{SS}={1\over p}[1+h_N(\{P({\bar e}_i{\bar S}^\perp)\})-h_{4N}(\{P(
{\bar s}_j^\perp{\bar e}_i{\bar S})\})].\label{finalSS}
\end{equation}

\section{relation of $Q_{SS}$ to quantum coherent information}
\label{sec4}

\begin{figure}
\epsfxsize=15cm
\leavevmode
\epsfbox{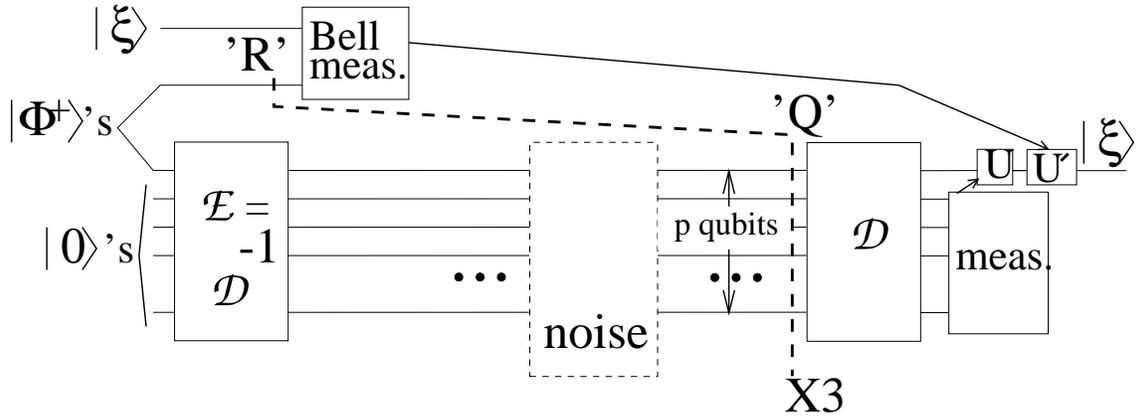}
\caption{The
Channel protocol as considered by \protect\cite{Schumacher2} in its treatment
of coherent information.  The `Q' subsystem in the one transmitted through
the channel, while the single qubit `R' remains behind.
\label{fig6}}
\end{figure}

The two noisy-channel transmission constructions which we have discussed
above are equivalent to yet a third one shown in Fig. \ref{fig6},
which has been extensively discussed in the literature
\cite{Schumacher1,Schumacher2,Barnum}.
The rationale of introducing the
one-qubit ancillary system `$R$' is that it is the minimum-size ancilla
required to ``purify'' the input of the channel, that is, to make it
part of a larger pure state \cite{Jozsa} (this is a different
sense of the word ``purification'' than used in \cite{purification}).  
In this scenario there is
an important information-theoretic measure, the {\em
coherent information}; at slice $X3$ this is given by the difference
of two von Neumann entropies:
\begin{equation}
I_e\equiv{1\over p}\left(S(\rho_Q)-S(\rho_{RQ})\right).
\end{equation}
Refs. \cite{Barnum,Alla} show that $I_e$ provides an
upper bound for the quantum channel capacity when maximized over all
possible input-state ensembles and quantum codes.  What we will show
is that the achievable Shor-Smolin capacity $Q_{SS}$ in fact {\em
attains} the coherent information for the same additive quantum code,
and for the input as in Fig. \ref{fig6}.  To establish this we need to
show the following two equalities:
\begin{equation}
S(\rho_Q)=1+h_N(\{P({\bar e}_i{\bar S}^\perp)\}),\label{sstoie1}
\end{equation}
and
\begin{equation}
S(\rho_{RQ})=h_{4N}(\{P({\bar s}_j^\perp{\bar e}_i{\bar S})\}).
\label{sstoie2}
\end{equation}
Establishing these just requires a consideration of how the noise acts
on the input state in Fig. \ref{fig6}.  For Eq.~(\ref{sstoie1}), we note that
the density matrix $\rho_Q$ {\em before} the action of the noise is
just an equal mixture of the $|0\rangle_{L0}$ and $|1\rangle_{L0}$
states, where the subscript 0 indicates that these vectors lie in the
eigenspace ${\cal C}_0$.  Each eigenspace ${\cal C}_i$, $0\le
i<2^{n-k}$, is spanned by a pair of vectors $|0\rangle_{Li}$,
$|1\rangle_{Li}$, where we can define the 0 and 1 vectors by
\begin{equation}
|0\rangle_{Li}={\bar e}_i|0\rangle_{L0},~
|1\rangle_{Li}={\bar e}_i|1\rangle_{L0},
\end{equation}
where ${\bar e}_i$ is the coset-generating operator (see
Fig. \ref{fig1}).  The importance of the basis $|0,1\rangle_{Li}$ is
that the density operator $\rho_{Q}$ {\em after} the action of the
depolarizing noise is diagonal in it.  The diagonal matrix elements
(i.e., the probabilities) for each vector is evaluated by noting that
the state $|0\rangle_{Li}$ is reached in four possible ways: 1) the
initial state is $|0\rangle_{L0}$ (with probability $\half$) and an
operator of the coset ${\bar e}_i{\bar S}$ is applied by the channel,
2) the initial state is $|0\rangle_{L0}$ and an operator of the coset
${\bar e}_i{\bar s}_z^\perp{\bar S}$ is applied by the channel, 3) the
initial state is $|1\rangle_{L0}$ (also with probability $\half$) and
an operator of the coset ${\bar e}_i{\bar s}_x^\perp{\bar S}$ is
applied by the channel, or 4) the initial state is $|1\rangle_{L0}$
and an operator of the coset ${\bar e}_i{\bar s}_y^\perp{\bar S}$ is
applied by the channel.  Each of these is given by the appropriate
weight polynomial, so
\begin{eqnarray}
\langle 0|\rho_Q|0\rangle_{Li}&=&
\half P({\bar e}_i{\bar S})+
\half P({\bar e}_i{\bar s}_z^\perp{\bar S})+
\half P({\bar e}_i{\bar s}_x^\perp{\bar S})+
\half P({\bar e}_i{\bar s}_y^\perp{\bar S})\\
&=&\half P({\bar e}_i{\bar S}^\perp).
\end{eqnarray}
The enumeration of the ways that the state $|1\rangle_{Li}$ can be
arrived at is identical, with 0s and 1s interchanged; so we find that
this matrix element is identical:
\begin{equation}
\langle 1|\rho_Q|1\rangle_{Li}=\langle 0|\rho_Q|0\rangle_{Li}.
\end{equation}
Because it is diagonal, the von Neumann entropy of $\rho_Q$ is just
the ordinary entropy of the probability distribution:
\begin{equation}
S(\rho_Q)=h_{2N}(\{\half P({\bar e}_i{\bar S}^\perp), 
\half P({\bar e}_i{\bar S}^\perp)\})=1+
h_N(\{P({\bar e}_i{\bar S}^\perp)\}).
\end{equation}
And thus Eq.~(\ref{sstoie1}) is established.  The reasoning needed to 
establish Eq.~(\ref{sstoie2}) is very similar: the joint state of systems
$R$ and $Q$ after encoding but before the noise is
\begin{equation}
\half|0_R\rangle|0_Q\rangle_{L0}+\half|1_R\rangle|1_Q\rangle_{L0}\equiv
\Phi^+_0.
\end{equation}
In this notation the $i$ in $\Phi^+_i$ means that the state in the $Q$
subsystem lies in the ${\cal C}_i$ eigenspace.  After the noise the
density matrix $\rho_{RQ}$ is diagonal in this generalized Bell basis,
with the probability of the state being $B_{ji}$ given by
\begin{equation}
\langle B_{ji}|\rho_{RQ}|B_{ji}\rangle=P({\bar e}_i{\bar s}_j^\perp
{\bar S}),\label{inter}
\end{equation}
since it is again only members of a particular coset that will produce
a final $B_{ji}$ state.  (This discussion can equivalently be given in
terms of the behavior of the $[p+1,k=0,d]$ code to which the composite
system belongs.) From Eq~(\ref{inter}), the desired result
Eq.~(\ref{sstoie2}) follows immediately, so the identity between the
Shor-Smolin capacity and the coherent information is established for
any code.

\section{$Q_{SS}$ for the ``cat'' code}
\label{sec5}
\subsection{Closed-form evaluation}
\label{closedform}

It has not proved easy to evaluate the Shor-Smolin capacity
Eq.~(\ref{finalSS}) (or the equivalent coherent information) for a
general concatenation.  But a closed-form evaluation has proved
possible for one important family of inner $[p,1,d]$ codes which we
refer to as ``cat'' codes.  In the cat code for $p\geq 2$ the
stabilizer group $\bar S$ is generated by the operators
\begin{equation}
\sigma_{z1}\sigma_{z2},~\sigma_{z1}\sigma_{z3},...,~\sigma_{z1}\sigma_{zp}.
\label{sigmazp}
\end{equation}
For this code the code space ${\cal C}_0$ is spanned by 
\begin{equation}
|0\rangle_{L0}=|\!\overbrace{000...}^{p\ {\rm qubits}}\rangle
\label{cat0}
\end{equation}
and
\begin{equation}
|1\rangle_{L0}=|111...\rangle.
\label{cat1}
\end{equation}
Thus, the source density matrix before passage through the channel is,
using the Schumacher-Nielsen notation (Fig. \ref{fig6}) \cite{Schumacher2},
\begin{equation}
\rho_{Q(in)}=\frac{1}{2}|000...\rangle\langle000...|+
       \frac{1}{2}|111...\rangle\langle111...|.
\end{equation}
A purification of this density matrix involving just one qubit in the
subsystem $R$ is
\begin{equation}
\Psi_{RQ}=\frac{1}{\sqrt{2}}(|0\!\overbrace{000...}^{p\ {\rm qubits}}\rangle+
|1\!\overbrace{111...}^{p\ {\rm qubits}}\rangle).
\end{equation}
Here the first qubit is the one belonging to system $R$.  This wavefunction
has been referred to as the ``cat state'' in the literature.

\begin{figure}
\epsfxsize=15cm
\leavevmode
\epsfbox{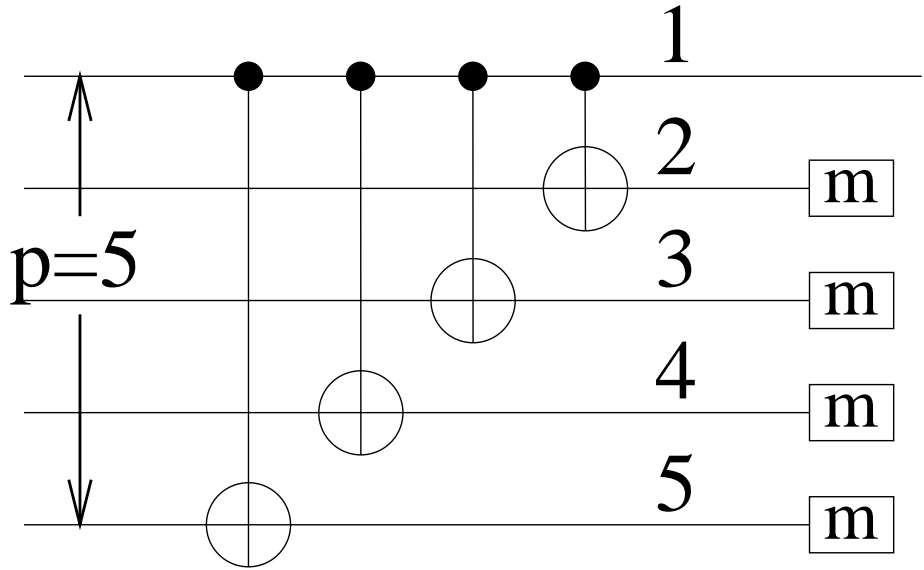}
\caption{Quantum network for decoding the ``cat'' code, shown for
$p=5$.  The same network is used for encoding.
\label{fig7}}
\end{figure}

The decoding network $\cal D$ for this code is extremely simple, just
consisting of the sequence of XOR gates shown in Fig. \ref{fig7}.
Shor and Smolin provide a detailed argument\cite{SSorig} for counting
all the probabilities in Eq.~(\ref{SSx}) by determining how each
different type of error process is modified by the XOR circuit.  We
summarize their results here: consider counting the probabilities of
the cases (including all members of one of the cosets of ${\bar
S}^\perp$) in which the measurements give exactly $r$ 1s, in
particular when the measurements of qubits 2 through $p-r$ give zero,
and qubits $p-r+1$ through $p$ give one.  It is obvious that the
counting is the same for any permutation of the qubits; this means
that there are $p-1 \choose r$ equivalent cosets being counted.  It is
this high multiplicity that permits the calculation to be tractable,
despite the fact that there are exponentially many (in $p$) coset
weight polynomials to be evaluated.

The further four subcases (i.e., the cosets of $\bar S$; see
Fig. \ref{fig1}) to be evaluated are:

1) The remaining qubits (qubit 1 of $Q$ and the qubit of $R$) are in
the state $\Psi^+$.  The error processes for which this occurs are
those where there are amplitude ($\sigma_x$) errors on qubits 1
through $p-r$, and an {\em even} number of phase ($\sigma_z$) errors
on any of the qubits.  We may forthwith calculate the probability of
this occurrence:
\begin{equation}
{\rm Pr}(\Psi^+,r)=\sum_{\rm t\ (even)}\sum_i{p-r\choose i}{r\choose t-i}
g^{p-r+t-i}f^{r-t+i}=2^{p-r-1}g^{p-r}(f+g)^r.
\end{equation}
Here $t$ is the total number of phase errors and $i$ is the number of
these phase errors occurring on the qubits which already have amplitude
errors (leading to a $\sigma_y$ error process).  The $t$ and $i$ sums
go over the full range for which the binomial coefficients are
non-zero.

2) The remaining state is $\Psi^-$.  For this the error processes are
those where there are amplitude ($\sigma_x$) errors on qubits 1
through $p-r$, and an {\em odd} number of phase ($\sigma_z$) errors on
any of the qubits.  In fact, it turns out that this count is exactly
the same as for $\Psi^+$:
\begin{equation}
{\rm Pr}(\Psi^-,r)=\sum_{\rm t\ (odd)}\sum_i{p-r\choose i}{r\choose t-i}
g^{p-r+t-i}f^{r-t+i}=2^{p-r-1}g^{p-r}(f+g)^r.
\end{equation}

3) The remaining state is $\Phi^+$.  In this case there must be
amplitude errors on qubits $p-r+1$ to $p$ (or no amplitude errors if
$r=0$), and there must be an even number of phase errors.  This gives
\begin{equation}
{\rm Pr}(\Phi^+,r)=\sum_{\rm t\ (even)}\sum_i{r\choose i}{p-r\choose t-i}
g^{r+t-i}f^{p-r-t+i}=\left\{\begin{array}{lr}\half\left((f+g)^p+
(f-g)^p\right),~&r=0,\\2^{r-1}g^r(f+g)^{p-r},&r>0.\end{array}\right.
\end{equation}

4) The remaining state is $\Phi^-$.  In this case there must be
amplitude errors on qubits $p-r+1$ to $p$, and there must be an odd
number of phase errors.  The result is the same as for $\Phi^+$ except
for the $r=0$ case:
\begin{equation}
{\rm Pr}(\Phi^-,r)=\sum_{\rm t\ (odd)}\sum_i{r\choose i}{p-r\choose t-i}
g^{r+t-i}f^{p-r-t+i}=\left\{\begin{array}{lr}\half\left((f+g)^p-
(f-g)^p\right),~&r=0,\\2^{r-1}g^r(f+g)^{p-r},&r>0.\end{array}\right.
\end{equation}
Plugging these expressions into Eqs.~(\ref{Prs}-\ref{QSSsimp}) permits
an efficient calculation of the Shor-Smolin capacity for the family
of cat codes.

The threshold of the cat-code family may be computed exactly for 
$p\rightarrow\infty$ using an asymptotic analysis.  Briefly, we
find that the capacity Eq.~(\ref{finalSS}) is dominated for large $p$ by
two contributions: 1) Those for the cosets of $\bar S$ with $r=0$ (recall
that $r$ is the number of ones in the measured syndrome).  We find
that this contribution goes like
\begin{equation}
Q_{SS}(r=0)=c\left(\frac{(f-g)^2}{f+g}\right)^p.
\end{equation}
2) Those for cosets with $r\approx{p\over 2}$.  This contribution has the
form
\begin{equation}
Q_{SS}(r\approx{p\over 2})=-\gamma(f)\left(\sqrt{8g(f+g)}\right)^p.
\end{equation}
Here $\gamma(f)>0$ is a fairly complicated function of $f$.  Nevertheless,
the threshold for $f$ specified by $Q_{SS}(f)=0$ is simply obtained by
equation the bases of these two contributions:
\begin{equation}
\frac{(f-g)^2}{f+g}=\sqrt{8g(f+g)}.
\label{infinity}
\end{equation}
The relevant root of this equation, $f\approx .81808$, is the asymptotic threshold.
We have not developed any simple intuitive understanding for why this
threshold should remain finite as $p\rightarrow\infty$, but nevertheless
remain worse than the threshold for finite $p$ as we will now see.

\subsection{Investigations of cat-code capacities}
\label{results}

\begin{figure}
\epsfxsize=15cm
\leavevmode
\epsfbox{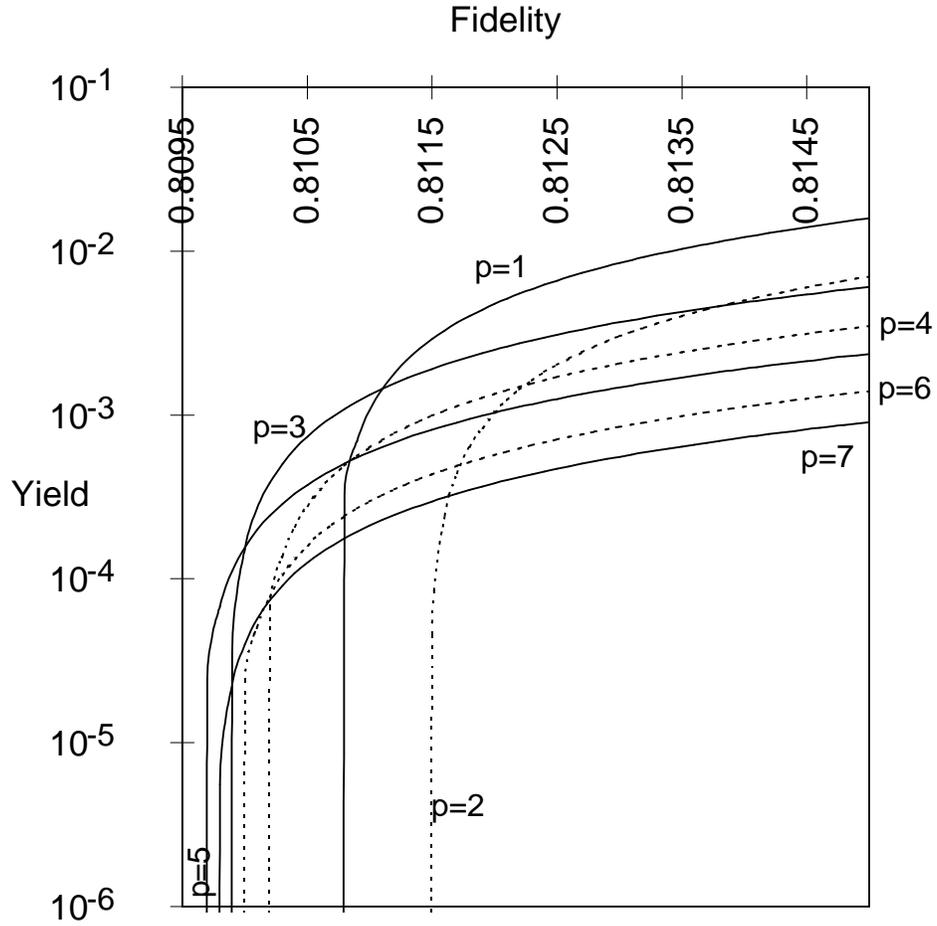}
\caption{The yield, i.e., capacity $Q_{SS}$, as a function of fidelity
$f$ for inner cat codes of size $p$ for of various values of $p$.
Note that the curves are all in $p$ order from $p=1$ to $p=7$ along
the right side of the graph.
\label{sschart}}
\end{figure}

The simplest codes to calculate are the cat codes
Eq. (\ref{cat0},\ref{cat1}).   The table shows the results for values of 
$p$ from one to fourteen.  The capacities $Q_{SS}$ of these codes
near $f=.81$ are shown in Fig. \ref{sschart}.  We note
that odd-$p$ codes work better than nearby even-$p$ codes;  the
lowest threshold fidelity in this family is achieved for $p=5$.

Generically, many other multiple-concatenation codes are possible
and may lead to better thresholds.
We explored the family of codes where the innermost code
has a rotated cat code for which the stabilizers are:
\begin{equation}
\sigma_{x1}\sigma_{x2},~\sigma_{x1}\sigma_{x3},...
\end{equation}
and the next-level code remains the ordinary cat code of
Eq. (\ref{sigmazp}).  The best code we found was for both inner and
outer cat-codes having $p=5$.  The capacity of this code was found to
be non-zero down to a fidelity of $f\approx .80944$, the best code
known\cite{footnote15}.  This threshold is still far above the best
known lower bound for the threshold of $f=\frac{3}{4}$
\cite{bruss,BDSW}.  Unfortunately, larger codes become computationally
intractable using our methods, because the number of distinct cosets
scales exponentially with $p$.  It is hoped that another approach,
perhaps an approximation method for coset weights, will permit a more
thorough exploration of concatenated codes.

\section{Conclusions}
\label{conclusions}

The obvious unanswered question which this work raises is, can any
finite capacity be achieved for even noisier depolarizing channels,
ones with $f$ below the lowest value, $.80944$, achievable with the
25-bit inner code, but above the absolute minimum threshold $f=0.75$
set by the no-cloning argument\cite{bruss}?  In other words, do there
exist even more clever non-random codes (recall \cite{nonad}) for
protecting qubits from high levels of noise?

It may be worthwhile to note here why we initially believed that the
use of inner codes of the cat type was a promising direction for
finding good codes for very noisy channels; this belief was based on
the property of degeneracy mentioned earlier.  While these motivations
may end up having no more than historical interest, since they have
not at present led us to any conclusive answer to the questions just
posed, we hope that it might assist some reader who is interested in
exploring these problems further.

{\em Degeneracy} is a property of quantum codes which has no analog
for classical error correcting codes.  Degeneracy arises from the fact
that two different error patterns can have indistinguishable effects
on a coded quantum state.  This is obviously impossible for a coded
binary (classical) string, but it is {\em obligatory} for additive
quantum codes; indeed, the cosets of $\bar S$ introduced in
Sec.~\ref{sec2} are precisely these groups of indistinguishable
errors.  A code is considered degenerate if some of the low-weight
($\leq\lfloor d/2\rfloor$ for an $[n,k,d]$ code) error patterns fall
in the same coset of $\bar S$ and are therefore indistinguishable.
The original 9-bit code of Shor\cite{SG} was degenerate; the 7-bit
code \cite{CS,Steane,Steane2} and the 5-bit code\cite{LF,BDSW} are
non-degenerate.

It is known \cite{ekert} that a Hamming-like bound could be easily
derived on the maximum attainable distance for a quantum code, {\em
provided} that it was non-degenerate.  However, in this work the
possibility remained open that degenerate codes could attain a greater
distance.  We were thus motivated to consider highly degenerate codes
for the attainment of high capacity, given the qualitative
relationship between code distance and capacity.  This possibility of
attaining large distance using degeneracy has subsequently been
rendered unlikely by a recent result of Rains\cite{Rains} who has
obtained a bound on $d$ which applies for both degenerate and
non-degenerate codes and which is tighter than the Hamming bound for a
substantial part of the $p,k,d$ parameter space.  Nevertheless the
fact is that the cat codes, which we have used successfully to attain
high capacity, are highly degenerate: single phase errors are all
indistinguishable, and all pairs of amplitude errors are
indistinguishable from the no-error process.  All this is true despite
the fact that the cat codes have very poor distance ($d=1$ for all 
$p$).

The best we can say about why this scheme succeeded is that the high
degeneracy, by making many outcomes indistinguishable, ``hides'' the
large amount of entropy which the very noisy channel adds to the
quantum states, thus permitting the average entropy $S_{X2}$ to be
below one over a greater range of $f$.  This reasoning is certainly
not rigorous; nevertheless, in an extensive Monte-Carlo
search of other additive codes, we found no other inner code with
$p\leq 5$ which does a better job than the cat code for reducing
the average entropy and hence attaining any higher capacity.  It
was further the observation that the cat code ``hides'' phase
error more effectively than amplitude error that motivated us
to consider a second level of concatenation, in which the innermost
code was a cat code with the role of amplitude and phase reversed.
Of course, this is what led us to the 25-bit code described above
which give the best capacity to date.

It is clear that further generalizations of this problem await
exploration.  The issue of attainable capacities for channels other
than the depolarizing channel is largely untouched.  It is fairly
clear that for the {\em generalized} depolarizing channel, in which
the error operators are still proportional to the Pauli matrices, but
with unequal probability amplitudes, the formalism developed here
(i.e., the weight polynomials, and the relation to coherent
information) will go through with little modification, so that would
be an easy direction for further study.  For the much larger space of
general channels, nothing better than our ``twirling'' arguments of
\cite{BDSW} (which bounds the capacity of any arbitrary
channel by that of a corresponding generalized depolarizing channel)
is presently known.  Further extensions of the formalism would
obviously also be desirable; a generalization of the present approach
for inner codes with $k>1$ would be desirable; also, asymptotic
expressions for the capacity which would not require an exact
evaluation of all the coset weight polynomials could lead to
significant progress.  Certainly there remains much to be done to
fully characterize the usefulness of the very noisy quantum channel.

We thank Charles H. Bennett, Artur Ekert, Daniel Gottesman, Emmanuel
Knill, John Preskill and Eric Rains for many helpful discussions.  We
are grateful to the Army Research Office for support.

\begin{table}
\begin{tabular}{lll|ll}
p&f&&p&f\\
\cline{1-5}
1&.81071&&9&.81002\\
2&.81148&&10&.81028\\
3&.80987&&11&.81032\\
4&.81010&&12&.81056\\
5&.80964\ \ \ Best&&13&.81062\\
6&.80991&&14&.81085\\
7&.80977&&$\cdots$&\\
8&.81004&&$\infty$&.81808\\
\end{tabular}
\caption{The value of the threshold fidelity $f$ for cat codes
of size  $p$.   Values of $p$ not 
shown all work less well than the random coding method ($p=1$).  
The value for $p=\infty$ is analytic from Eq. (\protect\ref{infinity}).
\label{table}}
\end{table}

\begin{figure}
\epsfxsize=15cm
\leavevmode
\epsfbox{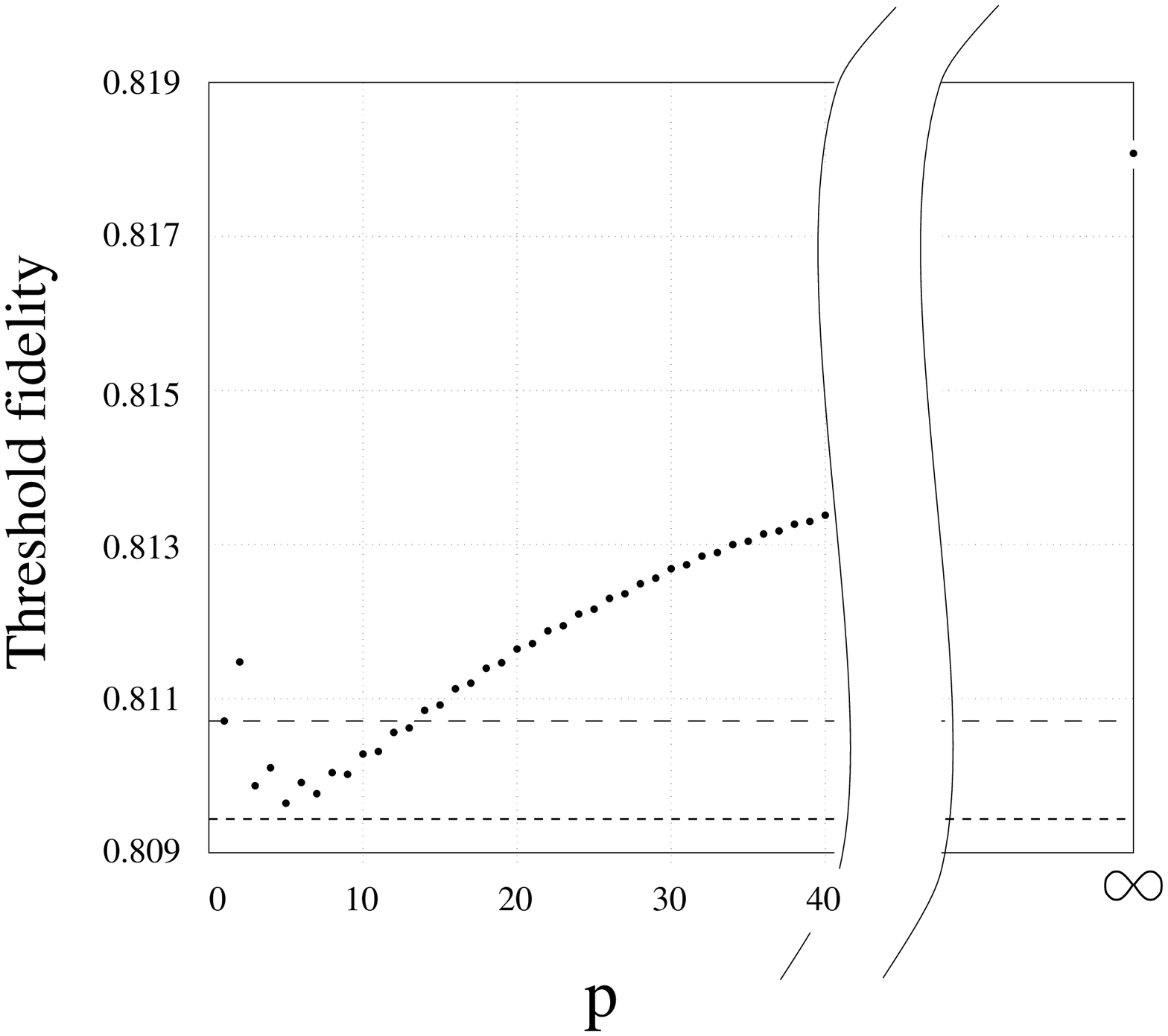}
\caption{The threshold value $f$ for which $Q_{SS}=0$ vs. $p$.  Note
that the points fall on two smooth curves, one for even $p$ and one
for odd $p$.  The value at $p\rightarrow\infty$, $f\approx .81808$,
is obtained by asymptotic analysis Eq. (\protect\ref{infinity}).  The
heavy dashed line at $f\approx .80944$ is the best known threshold,
for the twice concatenated 25-bit scheme
(Sec. \protect\ref{closedform}).  The light dashed line at $f\approx
.81071$ is the threshold for ordinary quantum random coding,
equivalent to the $p=1$ cat code.
\label{threshold}}
\end{figure}


\begin{references}
\bibitem{micro1} Quantum hash coding is a protocol for which a
vanishingly small fraction of errors are not corrected.  The distance
$d$ for a typical hash code can be obtained by the following counting
procedure: There are $2^{nS(f)}$ likely error patterns to be assigned
to $2^{n-k}$ possible hash strings.  It is highly likley that there
will be cases where two error patterns are at random assigned to the
same hash.  A smaller set of error patterns can be assigned uniquely
to hashes with a low probability of collision, provided that this
number is smaller than $\sqrt{2^{n-k}}$ (coming from the usual
birthday-paradox reasoning).  This fixes the distance between such
errors according to  
$2^\frac{n-k}{2}=2^{\frac{n}{2}S(f)}=2^{nS(1-\frac{d}{2n})}$.  
Thus the distance scales like the block size, being approximately 
$d\approx .8 (1-f)n$.
\bibitem{nonad} Of course, even more exotic codes exist, ones that do
not conform to the additive code classification at all.  The only
non-additive codes of which we are aware are in
\protect\cite{lafknill,leung,Hardin}.  We cannot rule out the
possibility that a non-additive code achieves the best possible
capacity for depolarizing channels.  Only the code in
\protect\cite{Hardin} is directly relevant the depolarizing channel
discussed in most of this paper.
\bibitem{footnote15} We did not search all possible cat-in-cat codes
up to 25 qubits, but the next best code we identified was
a $p=5$ inside a $p=3$ cat code which had a threshold at
$f\approx .80952$.  The $p=3$ inside a $p=5$ code was much
worse, having a threshold at $f\approx .81112$, worse even
than the regular hashing result for which $p=1$ and the threshold
is at $f\approx .81071$.
\end{references}
\end{document}